\documentclass[aip,amssymb,reprint,superscriptaddress]{revtex4-1}

\usepackage{mathtools} 
\usepackage{float}
\usepackage{epstopdf}
\usepackage{graphicx}
\usepackage{amsmath}
\usepackage{dcolumn}
\usepackage{bm}
\usepackage{bbold}
\usepackage{epstopdf}
\usepackage{amssymb}
\usepackage{mathrsfs}
\usepackage{amsfonts}
\usepackage{mathtools}
\usepackage{braket}
\usepackage{tabularx}
\usepackage{setspace}
\usepackage{dcolumn}

\usepackage{float}
\usepackage{array}
\newcolumntype{L}[1]{>{\raggedright\let\newline\\\arraybackslash\hspace{0pt}}m{#1}}
\newcolumntype{C}[1]{>{\centering\let\newline\\\arraybackslash\hspace{0pt}}m{#1}}
\newcolumntype{R}[1]{>{\raggedleft\let\newline\\\arraybackslash\hspace{0pt}}m{#1}}

\begin{document}

	\title[]{{Piezo-optomechanical coupling of a 3D microwave resonator to a bulk acoustic wave crystalline resonator}} 
	
	\author{N. C. Carvalho}
	\email[]{nataliaccar@gmail.com}
	\affiliation{Applied Physics Department, Gleb Wataghin Physics Institute, University of Campinas, 13083-859, Brazil.}
	\affiliation{ARC Centre of Excellence for Engineered Quantum Systems (EQuS), Department of Physics, 35 Stirling Hwy, 6009 Crawley, Western Australia.}
	
	\author{J. Bourhill}%
	\affiliation{ARC Centre of Excellence for Engineered Quantum Systems (EQuS), Department of Physics, 35 Stirling Hwy, 6009 Crawley, Western Australia.}
	
	\author{M. Goryachev}%
	\affiliation{ARC Centre of Excellence for Engineered Quantum Systems (EQuS), Department of Physics, 35 Stirling Hwy, 6009 Crawley, Western Australia.}
		
	\author{S. Galliou}
	\affiliation{FEMTO-ST Institute, Université Bourgogne Franche-Comté, CNRS, ENSMM, 25000 Besançon, France}
	
	\author{M. E. Tobar}
	\affiliation{ARC Centre of Excellence for Engineered Quantum Systems (EQuS), Department of Physics, 35 Stirling Hwy, 6009 Crawley, Western Australia.}%

	\date{\today}

	\begin{abstract} 
	
		We report the observation of coupling between a 3D microwave cavity mode and a bulk mechanical resonator mediated by piezoelectric and radiation pressure effects. The system is composed of a quartz bulk acoustic wave resonator placed inside a microwave re-entrant cavity, which is designed to act as both the electrodes for piezoelectric actuation as well as a 3D resonator. The cavity electromagnetic mode is modulated by a 5 MHz bulk acoustic wave shear mode, which is modeled and experimentally verified using the input-output formalism. Through finite element method simulations, we calculate the various contributions to the electromechanical coupling and discusss the potential of the system to reach high cooperativities as well as suitable applications.
	
	\end{abstract}

	\maketitle
	
	Nowadays, the pursuit of faster, more secure and efficient communication and information processing demands rigorous and diversified development of highly advanced technologies. To this purpose, hybrid systems have been intensely researched, as they combine the advantages of different platforms while avoiding their specific drawbacks. In particular, microwave devices figure as a rich means of  photon-phonon coupling, which is a multifold platform for the investigation of fundamental science and applied physical systems. In this sense, electrostriction has been explored using piezoelectric photonic crystals to couple microwave and optical photons \cite{bochmann2013, fong2014}, showing that piezomechanics associated to optomechanics is a promising path for the implementation of microwave to optical (MO) interconversion -  a convenient route to take advantage of distinct wavelengths in order to promote the optimized information transfer required for modern communication.
	
	In this field of cavity optomechanics, device architectures have been diversified with structures such as membrane-in-the-middle cavities \cite{thompson2008}, whispering gallery mode resonators \cite{schliesser2010,bourhill2015} and photonic crystals \cite{cotrufo2018} demonstrating numerous breakthrough results, such as gravitational wave detection \cite{abbott2016}, tests of quantum gravity\cite{bourhill2015,bushevQG}, ground state cooling \cite{teufel2011} and optomechanically induced transparency \cite{weis2010}. 
	
	In parallel, Renning et al. \cite{renninger2018} demonstrated valuable features of bulk acoustic wave (BAW) resonators for several quantum enabled experiments, with a study of high-coherence phonons driven by optical fields, opening new possibilities for the investigation of quantum mechanics using mesoscopic systems, precision measurements and high-fidelity information processing. BAW devices are phonon-confining structures and an acoustic analog to a Fabry-Perot cavity \cite{goryachev2014e}, where phonons propagating in a plane are reflected at the boundary of the resonator and the external medium. Such phononic cavities are recognized by their potential to exhibit extremely high mechanical quality factors, of greater than billions at cryogenic temperatures, which is a desirable characteristic for experiments demanding large coherence times \cite{galliou2013}. 

	\begin{figure}[!htb]
		\begin{center}
			\includegraphics[width=0.75\columnwidth]{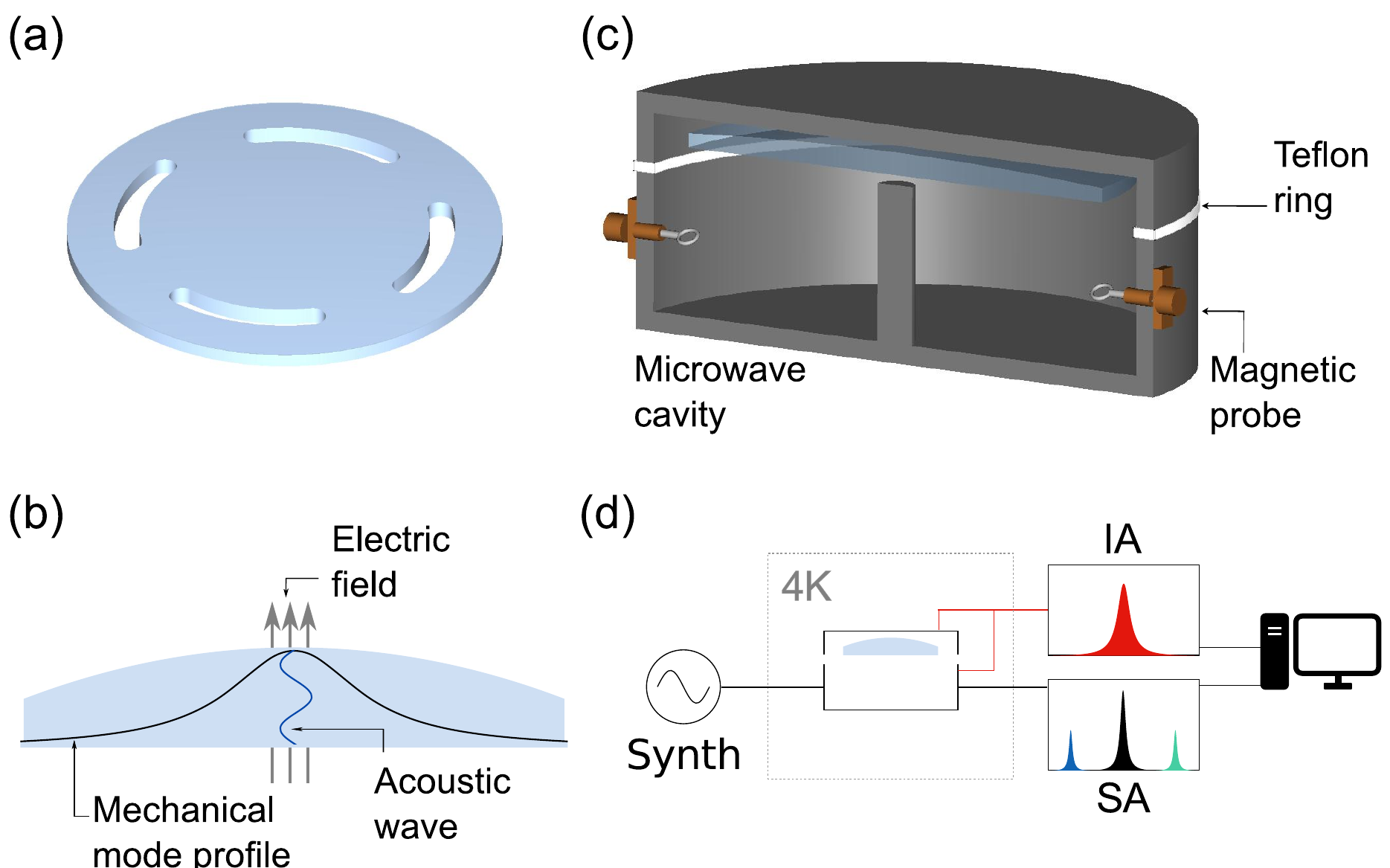}
			\caption{(a) BAW resonator. (b) Thickness modes driven by an electric field and forming standing shear acoustic waves. (c) Illustration of the MC-BAW system. (d) Experimental setup for microwave field modulation (Synth: RF syntesizer; IA: Impedance analyzer; SA: Spectrum analyzer). 
				\label{fig1}}
		\end{center}
	\end{figure}

	In this paper, we explore the piezoelectricity of a quartz BAW resonator placed into a 3D microwave cavity (MC) in order to achieve coupling between ultra-high-quality-factor-phonons and photons. In the discussed system, the BAW device plays the role of the mechanical resonator and can be excited via piezoelectricity as well as radiation pressure optomechanics. We report experimental results analysed through finite element method (FEM) simulations, which agree with the theoretical model proposed. This investigation provides important information concerning the nature of the coupling of the bulk phonons to the electromagnetic MC modes, demonstrating the potential of the device to be implemented in quantum enabled applications and fundamental physics.	
		
	\begin{figure}[!htb]
		\begin{center}
			\includegraphics[width=0.6\columnwidth]{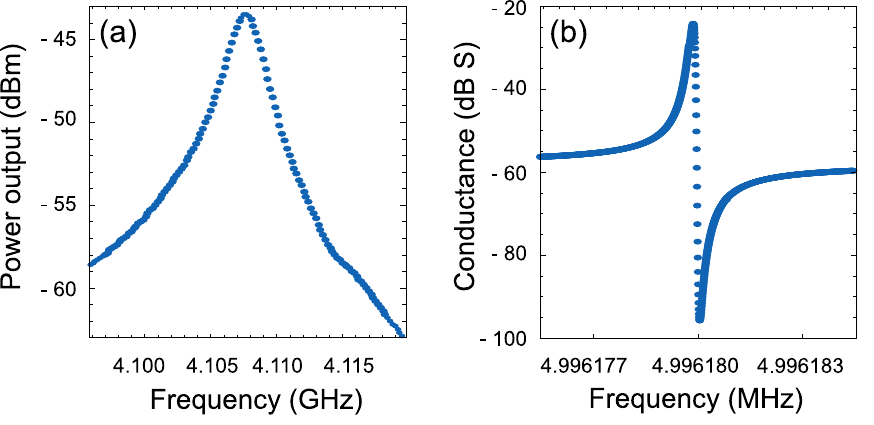}
			\caption{Measured  (a) microwave cavity resonance and (b) bulk acoustic resonance. 
				\label{fig2}}
		\end{center}
	\end{figure}
		
	The BAW resonator is made from pure quartz and is shown in Fig. \ref{fig1} (a). It has a plano-convex geometry, exaggerated in Fig. \ref{fig1} (b). A bulk crystalline system with such a design has the key feature of operating in very low-loss regimes when tuned to their thickness modes. Such resonances have a Hermitian-Gaussian distribution due the resonator{'}s curved shape, avoiding energy leakage through the clamping points. The MC-BAW set up is shown in Fig. \ref{fig1} (c). The re-entrant microwave mode generated in this cavity geometry has a strong axial electric field in the cavity gap (narrow volume between the top of the cavity central post and the opposite wall)  \cite{lefloch2013}. This is an optimal configuration as the field overlaps with the maximum mechanical displacement, improving the interaction between the MC and the BAW mode. 
	
	The MC is made of niobium and, at the operating temperature of 4 K, resonates at 4.1 GHz with intrinsic linewidth of 2 MHz. The Lorentzian peak obtained through a transmission measurement is shown in Fig. \ref{fig2} (a). The BAW resonator can exhibit three types of thickness vibration resonances: longitudinal (A-modes), fast shear (B-modes) and slow shear (C-modes) \cite{goryachev2014e}. In the experiment, a 5 MHz BAW mode was investigated. It is the third overtone C-mode with Q-factor measured as approximately 20 million at cryogenic temperatures. Fig.  \ref{fig2} (b) shows the measured mechanical mode. To excite the acoustic resonances, the cavity top and bottom structures were  isolated from each other by a Teflon ring, so they could function as electrodes and used to actuate the BAW resonator piezoelectrically. However, such a design element prevents the cavity from reaching the superconducting regime. The use of contactless electrodes, though, is an important characteristic of the system, as it reduces the coupling of the mechanical resonator to the environment, contributing to a more effective phonon-trapping. 
	
	The MC-BAW system dynamics can be described by the following Hamiltonian:
	
	\begin{multline} \label{eq1}
		H = \hbar \omega_{cav} a^{\dagger}a + \hbar \Omega_{m} b^{\dagger}b +\hbar g_0^{rp} (a^{\dagger}a)(b^{\dagger} + b) + \\
		\hbar g_0^{pzt}(c^{\dagger} + c)(b^{\dagger} + b),
	\end{multline}
	
	\noindent where $a~(a^{\dagger})$, $b~(b^{\dagger})$ and $c~(c^{\dagger})$, are the bosonic 	annihilation (creation) operators of the MC, the acoustic modes and the electrodes driving field, respectively; $\omega_{cav}$ and $\Omega_m$ are the MC and BAW resonance frequencies.
		
	The optomechanical and piezomechanical coupling rates are given by $g_0^{rp}$ and $g_0^{pzt}$. The former originates from radiation pressure and must be decomposed as $g_0^{rp}$ = $g_0^{mb}$ + $g_0^{pe}$, where $g_0^{mb}$ corresponds to the thickness movement of the acoustic resonator, which modifies the boundary conditions of the MC field; and $g_0^{pe}$ relating to the strain-induced change in the refractive index of the BAW resonator, known as the photoelastic effect. For the second interaction involving $g_0^{pzt}$, the inverse piezoelectric effect produces a deformation of the atomic quartz lattice, also causing a modulation of the refractive index. However, differently from the optomechanical effect, piezoelectricity has a linear relationship with respect to the electric field, as explicitly manifested in Eq. (\ref{eq1}).
	
	The three forms of photon-phonon coupling can be calculate by the following overlap field integrals \cite{espinel2017, zou2016}:
		
	\begin{equation} \label{eq2}
		g_{mb} = - \frac{\omega_{cav}}{2}  \oint_{S} (\textbf{u} \cdot \hat{n}) (\delta \epsilon_{mb}|\textbf{E}|^2 - \delta \epsilon_{mb}^{-1}|\textbf{D}|^2)  \hspace{3pt} dA,
	\end{equation}	
	\begin{equation}\label{eq3}
		g_{pe} =  - \frac{\omega_{cav}}{2}\int_{V} \textbf{E}^{*} \cdot \delta \epsilon_{pe} \cdot \textbf{E} \hspace{3pt} dV,
	\end{equation}
	\begin{equation}\label{eq4}
		g_{pzt} = \frac{\omega_{cav}}{2} \int_{V} (\textbf{S}^*  \cdot e_{pzt}^\textit{T}  \cdot \textbf{E} + \textbf{E}^* \cdot e_{pzt}  \cdot \textbf{S} \hspace{3pt}) dV,
	\end{equation}
		
	\noindent where $g_{j} = g_0^{j}/x_{zpf}$ with $j=\{mb,pe,pzt\}$,
	$x_{zpf}=\sqrt{\hbar/2m_\text{eff}\Omega_m}$ is the zero point fluctuations and $m_\text{eff}$ is the effective modal mass of the mechanical resonator. Also, $\textbf{u}$  and $\hat{n}$  are the mechanical displacement and the unitary normal vectors, $\delta \epsilon_{mb} = \epsilon_{0}(n_1^2 - n_2^2)$ and $\delta \epsilon_{mb}^{-1} = \epsilon_{0}^{-1}(n_1^{-2} - n_2^{-2})$, with $n_1$ and $n_2$ being the refractive indices of dielectric (in this case quartz) and air, respectively. The electric, electric displacement and strain fields are given by $\textbf{D}$, $\textbf{E}$ and $\textbf{S}$ (complex conjugates: $\textbf{E}^*$ and $\textbf{S}^*$), $\delta_{pe}$ stands for the permittivity perturbation tensor, which is the scalar product of the dielectric photoelastic and strain tensors, and $e_{pzt}$ is the piezoelectric coupling matrix (with transpose $e_{pzt}^{\textit{T}}$). In this system, the rate equations of the  optical and mechanical amplitudes are:
	
	\begin{equation} \label{eq5}
		\dot{a} = -(i\Delta + \frac{\kappa_i}{2})a -ig_0^{rp}a(b+b^{\dagger})-\sqrt{\frac{\kappa_{ex}}{2}}\hspace{2pt}a_{in},
	\end{equation}	
	
	\begin{equation} \label{eq6}
		\dot{b} = -(i\Omega_m + \frac{\gamma_i}{2})b -ig_0^{rp}a^{\dagger}a-ig_0^{pzt}(c+c^{\dagger})-\sqrt{\frac{\gamma_{ex}}{2}}\hspace{2pt}b_{in},
	\end{equation}
	
	\noindent with the intrinsic microwave cavity (acoustic) linewidth given by  $\kappa_{i}$ ($\gamma_{i}$) and the extrinsic microwave (acoustic) intensity decay rate $\kappa_{ex} = \kappa - \kappa_i$ ($\gamma_{ex}$). 
	
	It is convenient to separate the effects acting on the acoustic resonator with regard to their origin. Therefore, there are forces due to radiation pressure, $F_{rp}$, piezoelectricity, $F_{pzt}$, and thermal fluctuations, $F_{th}$, all summing up to induce mechanical displacements. It is assumed that the latter is small compared to the former two forcing terms such that it can be ignored. Even though only the optomechanical interaction is able to provide the nonlinearity necessary for modulation and feedback between the acoustic and microwave resonators, it is possible to demonstrate that the piezoelectric coupling can enhance the modulation of the photonic field. 
	
	If the acoustic pump is set on the mechanical resonance frequency, in the resolved sideband region \cite{balram2016} it can be shown that the sideband amplitudes are given by:
	
	\begin{equation}\label{eq7}
		|\alpha_{\pm}|^2=\frac{4|\alpha_0|^2 |\xi|^2 (g_0^{rp})^2 (g_0^{pzt})^2}{(4|\alpha_0|^2(g_0^{rp})^2+\gamma_i\kappa/2)^2+\gamma_i^2(\Delta \pm \Omega_m)^2},
	\end{equation}
		
	\noindent where $\alpha_0$ is the steady-state amplitude of the electromagnetic drive, $\xi$ the amplitude of the acoustic drive and $\Delta $ stands for the detuning of the electromagnetic pump from the MC resonance frequency.
		
	Here, it{'}s clear that the response contains combination of radiation pressure and piezoelectric coupling. Thus, a large piezoelectric drive heightens the intensity of the cavity field modulation. The phase relationship between the microwave and acoustic driving fields is also relevant, as it could result in suppression of the mechanical displacement instead of amplification. Interesting outcomes have been observed exploring such destructive and constructive interferences\cite{balram2016}. In this work, however, the driving fields are connected to the same phase reference, therefore such considerations are not necessary. 
	
	\begin{figure}[!htb]
		\begin{center}
			\includegraphics[width=0.75\columnwidth]{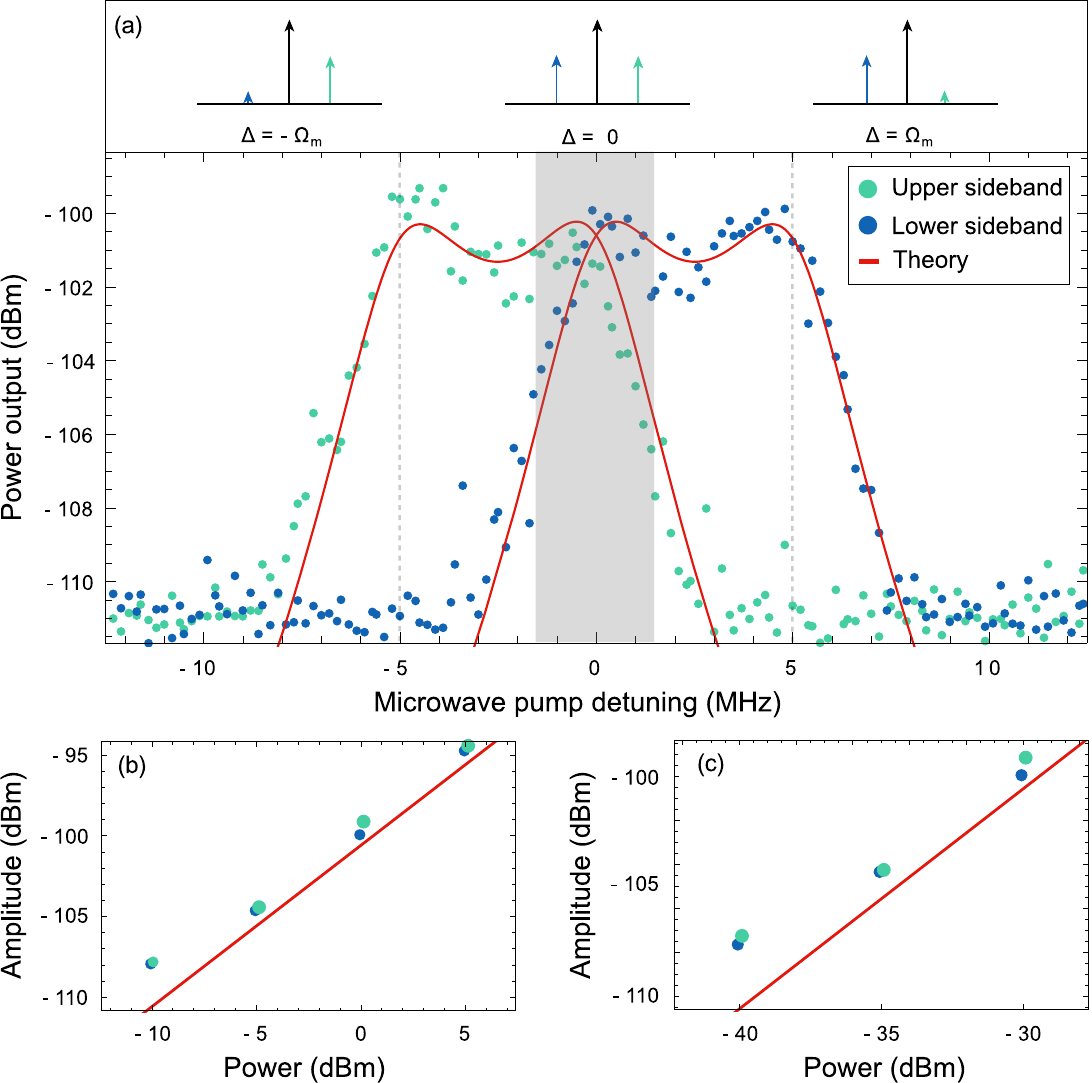}
			\caption{(a) Top: scheme of the sidebands induced on the transmitted microwave cavity signal. Bottom: measured sideband asymmetry with theoretical fit according to Eq. \ref{eq7}. Dashed lines: $\pm$ 5 MHz frequency shift. Shaded area: microwave cavity linewidth. Peak amplitude of the sidebands as a function of the (b) RF synthesizer and (c) impedance analyzer output power. 
				\label{fig3}}
		\end{center}
	\end{figure}
	
	The electrodes were connected to a vector network analyzer set for impedance analysis. This drove and monitored the acoustic resonance in reflection. Using the scheme in Fig. \ref{fig1} (d), the cavity electromagnetic field was modulated by the BAW 5 MHz-resonance, generating sidebands at the frequencies $\omega_{cav} \pm \Omega_m$. The microwave photonic field was driven by a two-port system through magnetic loop probes. Thus, the cavity was pumped by a microwave frequency synthesizer, with read out performed by a spectrum analyzer. All the equipment was connected to a frequency reference hydrogen maser and the MC-BAW system was cooled down to cryogenic temperatures in a pulse-tube cryocooler with internal temperature stabilised at 4 K.	
	
	Measurements were undertaken to observe the response of the piezo-optomechanical interaction as the MC pump was detuned from the resonance. Fig. \ref{fig3} (a) displays the data demonstrating higher sideband amplitudes when the pump is tuned on resonance or when the detuning matched the mechanical mode frequency. Fig. \ref{fig3} (b) shows the amplitude of the sidebands when the synthesizer is tuned to the microwave resonance and its input power swept. Analogously, Fig. \ref{fig3} (c) shows the amplitude of the sidebands as a function of the power applied to the electrodes, presenting the same linear behaviour as predicted by Eq. \ref{eq7}. 
	
\begin{figure}[!htb]
		\begin{center}
			\includegraphics[width=0.72\columnwidth]{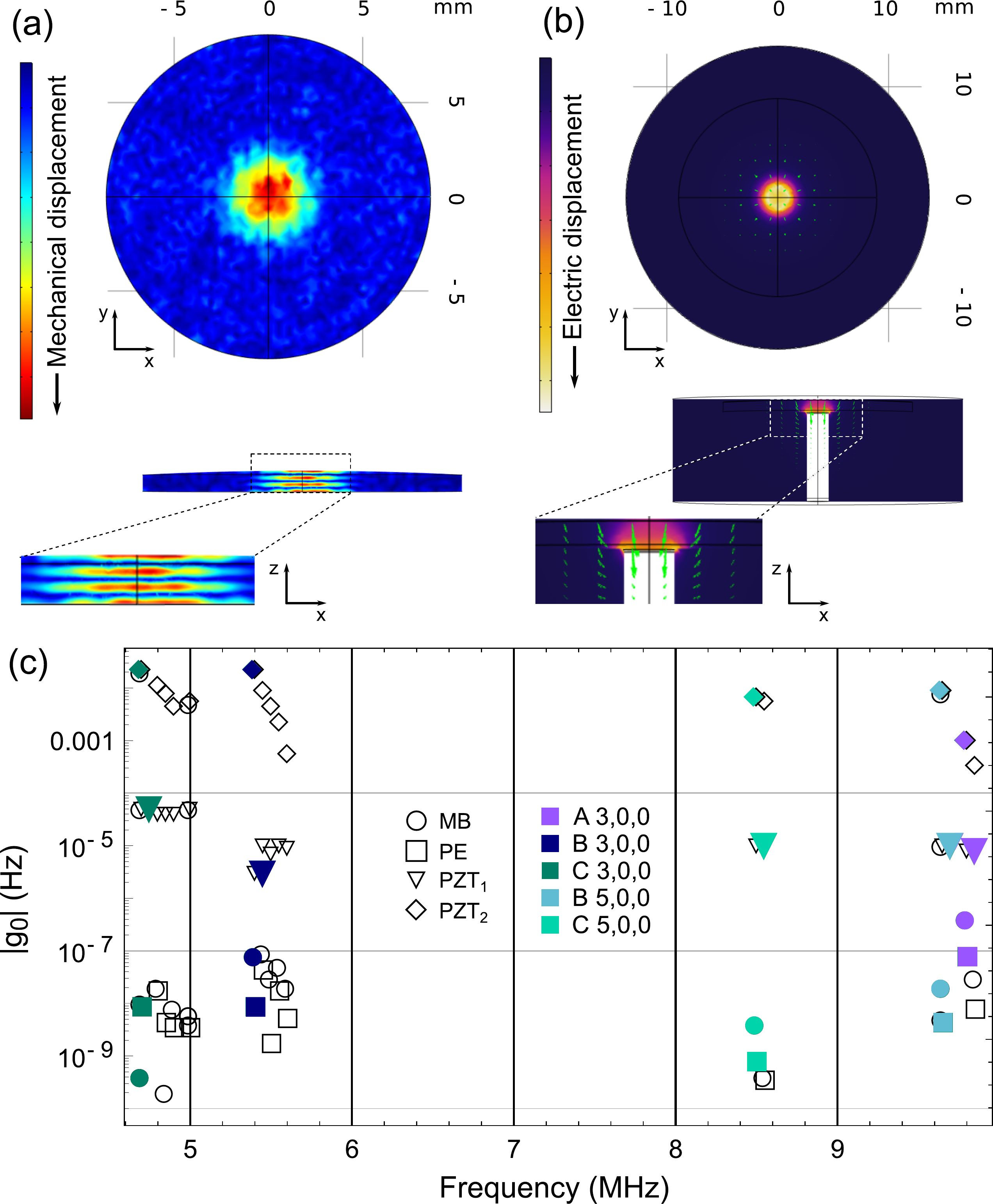}
			\caption{FEM model of the (a) third overtone C acoustic mode (C$_{3,0,0}$) and (b)  microwave re-entrant cavity mode. (Bottom: transverse planes and magnified views. Green arrows: proportional electric displacement field). (c) Vacuum optomechanical and piezomechanical coupling strengths of the A, B and C thickness modes (MB: moving boundaries; PE: photoelastic; PZT$_1$: piezoelectric with electrodes drive; PZT$_2$: piezoelectric with cavity field drive).
				\label{fig4}}
		\end{center}
	\end{figure}

	In order to identify the physical nature of the coupling between the electromagnetic mode and the mechanical resonance a FEM model was built and used to evaluate $g_{mb}$, $g_{pe}$ and $g_{pzt}$ through Eqs. (\ref{eq2})-(\ref{eq4}).  Figs. \ref{fig4} (a) and (b) present the simulated MC field and the 5 MHz BAW mode. 
	
	As the acoustic resonator is a doubly-rotated stress compensated cut of pure quartz, material tensors give rise to the quasi-shear vibrational modes. Hence, such asymmetric geometries can only be reproduced by a 3D model. Tridimensional simulations, though, are very computationally expensive and limit the quality of the FEM mesh due to memory and time consumption. Despite this, the model is still suitable to estimate the order of magnitude of the piezo-optomechanical coupling. 
	
	The piezoelectric component was calculated for two arrangements: first, when the acoustic mode is excited by the electrodes and then when it is resonant with the MC field. Only the first case was experimentally investigated. The second, could be explored using the beating of two pumps detuned from each other by the BAW mode frequency. Fig. 4 (c) shows the vacuum coupling strength due to optomechanical and piezoelectric effects in a frequency range that encompass a series of other shear and longitudinal acoustic modes. Many of these acoustic resonances were observed experimentally in the mechanical spectrum, and some of them were characterized by \cite{galliou2013}. The simulated third overtone C-mode is located at approximately 4.7 MHz (6\% deviated from the measured frequency), with calculated piezomechanical coupling rate of $g_0^{pzt}=4 \times 10^{-5}$ Hz for the case of electrode driving, which corresponds with the experimental data (see red curves in Fig. \ref{fig3}). 
	
	In the piezoelectric coupling rate shown in Fig. \ref{fig4} (c), the subscript 1 stands for electrical actuation and 2 for cavity field drive. These results indicate that, for the third overtone C-mode, the radiation pressure coupling is dominated by the photoelastic effect, although this is not a regular trend for the other modes. With respect to the piezolectric effect, coupling resulting from resonantly excited acoustic modes using the microwave cavity field is, in general, three orders of magnitude larger than the excitation with electrodes.
	
	In conclusion, FEM indicates we were able to reach an effective piezo-optomechanical coupling rate, $g^\text{eff}_0 = \sqrt{g_0^{pe} g_0^{pzt1}}$, of the order of $\mu Hz$, which agrees with the theory used to fit the data. We can show that even with modest coupling strengths, such devices can achieve reasonable cooperativity values with simple optimization. Simulations demonstrate that making the cavity field resonant with the BAW mode - using a pump and probe, or two-tone configuration - we can piezoelectrically drive the crystal phonons more efficiently (Fig. \ref{fig4} (c)). The fact that it is possible to excite acoustic modes by replacing the electrodes by a two-tone microwave pump, gives yet another advantage: the Teflon spacer can be removed. This would permit the niobium cavity to become fully superconducting. Niobium made cavities have reported electrical Q-factors \cite{bassan2008} as high as $10^8$. Then, using Eq. \ref{eq7} the piezo-optomechanical cooperativity can be approximated as:
	
	\begin{equation}\label{eq8}
		C_{om\_pzt} = \frac{4 (g^\text{eff}_0)^2 n_{cav}}{\kappa \gamma} 
	\end{equation}
		
	\noindent where $\gamma$ is the total mechanical linewidth and $n_{cav}$ is the number of photons inside the cavity. A microwave cavity field driven at $0.1$ $\mu W$, operating with a Q-factor of $10^8$ and $g^\text{eff}_0 = \sqrt{g_0^{pe} g_0^{pzt2}} \sim 10^{-5}$ Hz, could theoretically achieve $C_{om\_pzt}>10^4$. 
	
	These levels of cooperativity point to an ability to produce dressed states in the system, and reveal its potential to be employed in a series of applications. Zou et al. \cite{zou2016} proposed a similar device, where an AlN-on-Si chip is placed in a microwave re-entrant cavity to enable coherent M-O conversion. In the same direction, Vainsencher \cite{vainsencher2016} and Javerzac-Galy \cite{javerzac2016},  have investigated electro-optomechanical devices for use as quantum transducers. Moreover, Han et al. \cite{han2016} have specifically worked with bulk acoustic resonators, reaching electro-mechanical strong coupling, which is also valuable for quantum memory implementions.
		
	Therefore, the MC-BAW system is a feasible platform for hybrid quantum devices, potentially enabling quantum state transfer, coupling with superconducting qubits and tests of fundamental theories \cite{carvalho2016}. In this sense, this piezo-optomechanical device is favored by its high mechanical frequency-quality-factor product, high power handling and large effective modal mass. Also, it supports higher frequency acoustic modes, that can allow direct ground state cooling via a dilution refrigerator. Acoustic frequencies approaching 1 GHz have already been measured in a quartz BAW resonator \cite{goryachev2013o}. Additionally, other materials with strong photoelastic and piezoelectric response can be shaped as a confocal resonator and be explored in the same way.
	
	In summary, we have designed and tested a device able to sustain piezo-optomemechanical coupling between a microwave 3D cavity and a BAW resonator. This hybrid device was also theoretically and numerically modelled, being thoroughly investigated with regard to the nature of the different coupling mechanisms, namely: piezomechanics and optomechanics. It was concluded that the system can reach cooperativities as high as $10^4$ with realistic adjustments, demonstrating its potential to integrate systems with applications in MO wavelength conversion and investigation of nonclassical effects at the macroscopic scale. 
	
	We acknowledge fruitful discussions with R. Benevides and T. P. M. Alegre. This research was supported by the Australian Research Council, CE170100009 and by the Conselho Nacional de Desenvolvimento Científico e Tecnológico (CNPq - Brazil).


\begin{thebibliography}{23}%
		\makeatletter
		\providecommand \@ifxundefined [1]{%
			\@ifx{#1\undefined}
		}%
		\providecommand \@ifnum [1]{%
			\ifnum #1\expandafter \@firstoftwo
			\else \expandafter \@secondoftwo
			\fi
		}%
		\providecommand \@ifx [1]{%
			\ifx #1\expandafter \@firstoftwo
			\else \expandafter \@secondoftwo
			\fi
		}%
		\providecommand \natexlab [1]{#1}%
		\providecommand \enquote  [1]{``#1''}%
		\providecommand \bibnamefont  [1]{#1}%
		\providecommand \bibfnamefont [1]{#1}%
		\providecommand \citenamefont [1]{#1}%
		\providecommand \href@noop [0]{\@secondoftwo}%
		\providecommand \href [0]{\begingroup \@sanitize@url \@href}%
		\providecommand \@href[1]{\@@startlink{#1}\@@href}%
		\providecommand \@@href[1]{\endgroup#1\@@endlink}%
		\providecommand \@sanitize@url [0]{\catcode `\\12\catcode `\$12\catcode
			`\&12\catcode `\#12\catcode `\^12\catcode `\_12\catcode `\%12\relax}%
		\providecommand \@@startlink[1]{}%
		\providecommand \@@endlink[0]{}%
		\providecommand \url  [0]{\begingroup\@sanitize@url \@url }%
		\providecommand \@url [1]{\endgroup\@href {#1}{\urlprefix }}%
		\providecommand \urlprefix  [0]{URL }%
		\providecommand \Eprint [0]{\href }%
		\providecommand \doibase [0]{http://dx.doi.org/}%
		\providecommand \selectlanguage [0]{\@gobble}%
		\providecommand \bibinfo  [0]{\@secondoftwo}%
		\providecommand \bibfield  [0]{\@secondoftwo}%
		\providecommand \translation [1]{[#1]}%
		\providecommand \BibitemOpen [0]{}%
		\providecommand \bibitemStop [0]{}%
		\providecommand \bibitemNoStop [0]{.\EOS\space}%
		\providecommand \EOS [0]{\spacefactor3000\relax}%
		\providecommand \BibitemShut  [1]{\csname bibitem#1\endcsname}%
		\let\auto@bib@innerbib\@empty
		\bibitem [{\citenamefont {Bochmann}\ \emph {et~al.}(2013)\citenamefont
			{Bochmann}, \citenamefont {Vainsencher}, \citenamefont {Awschalom},\ and\
			\citenamefont {Cleland}}]{bochmann2013}%
		\BibitemOpen
		\bibfield  {author} {\bibinfo {author} {\bibfnamefont {J.}~\bibnamefont
				{Bochmann}}, \bibinfo {author} {\bibfnamefont {A.}~\bibnamefont
				{Vainsencher}}, \bibinfo {author} {\bibfnamefont {D.~D.}\ \bibnamefont
				{Awschalom}}, \ and\ \bibinfo {author} {\bibfnamefont {A.~N.}\ \bibnamefont
				{Cleland}},\ }\bibfield  {title} {\enquote {\bibinfo {title} {Nanomechanical
					coupling between microwave and optical photons},}\ }\href@noop {} {\bibfield
			{journal} {\bibinfo  {journal} {Nature Physics}\ }\textbf {\bibinfo {volume}
				{9}},\ \bibinfo {pages} {712--716} (\bibinfo {year} {2013})}\BibitemShut
		{NoStop}%
		\bibitem [{\citenamefont {Fong}\ \emph {et~al.}(2014)\citenamefont {Fong},
			\citenamefont {Fan}, \citenamefont {Jiang}, \citenamefont {Han},\ and\
			\citenamefont {Tang}}]{fong2014}%
		\BibitemOpen
		\bibfield  {author} {\bibinfo {author} {\bibfnamefont {K.~Y.}\ \bibnamefont
				{Fong}}, \bibinfo {author} {\bibfnamefont {L.}~\bibnamefont {Fan}}, \bibinfo
			{author} {\bibfnamefont {L.}~\bibnamefont {Jiang}}, \bibinfo {author}
			{\bibfnamefont {X.}~\bibnamefont {Han}}, \ and\ \bibinfo {author}
			{\bibfnamefont {H.~X.}\ \bibnamefont {Tang}},\ }\bibfield  {title} {\enquote
			{\bibinfo {title} {Microwave-assisted coherent and nonlinear control in
					cavity piezo-optomechanical systems},}\ }\href@noop {} {\bibfield  {journal}
			{\bibinfo  {journal} {Physical Review A}\ }\textbf {\bibinfo {volume} {90}},\
			\bibinfo {pages} {051801} (\bibinfo {year} {2014})}\BibitemShut {NoStop}%
		\bibitem [{\citenamefont {Thompson}\ \emph {et~al.}(2008)\citenamefont
			{Thompson}, \citenamefont {Zwickl}, \citenamefont {Jayich}, \citenamefont
			{Marquardt}, \citenamefont {Girvin},\ and\ \citenamefont
			{Harris}}]{thompson2008}%
		\BibitemOpen
		\bibfield  {author} {\bibinfo {author} {\bibfnamefont {J.~D.}\ \bibnamefont
				{Thompson}}, \bibinfo {author} {\bibfnamefont {B.~M.}\ \bibnamefont
				{Zwickl}}, \bibinfo {author} {\bibfnamefont {A.~M.}\ \bibnamefont {Jayich}},
			\bibinfo {author} {\bibfnamefont {F.}~\bibnamefont {Marquardt}}, \bibinfo
			{author} {\bibfnamefont {S.~M.}\ \bibnamefont {Girvin}}, \ and\ \bibinfo
			{author} {\bibfnamefont {J.~G.~E.}\ \bibnamefont {Harris}},\ }\bibfield
		{title} {\enquote {\bibinfo {title} {Strong dispersive coupling of a
					high-finesse cavity to a micromechanical membrane},}\ }\href@noop {}
		{\bibfield  {journal} {\bibinfo  {journal} {Nature}\ }\textbf {\bibinfo
				{volume} {452}},\ \bibinfo {pages} {72--75} (\bibinfo {year}
			{2008})}\BibitemShut {NoStop}%
		\bibitem [{\citenamefont {Schliesser}\ and\ \citenamefont
			{Kippenberg}(2010)}]{schliesser2010}%
		\BibitemOpen
		\bibfield  {author} {\bibinfo {author} {\bibfnamefont {A.}~\bibnamefont
				{Schliesser}}\ and\ \bibinfo {author} {\bibfnamefont {T.~J.}\ \bibnamefont
				{Kippenberg}},\ }\href {www.scopus.com} {\emph {\bibinfo {title} {Cavity
					optomechanics with whispering-gallery mode optical micro-resonators}}},\
		\bibinfo {series} {Advances in Atomic, Molecular and Optical Physics},
		Vol.~\bibinfo {volume} {58}\ (\bibinfo {year} {2010})\ pp.\ \bibinfo {pages}
		{207--323}\BibitemShut {NoStop}%
		\bibitem [{\citenamefont {Bourhill}, \citenamefont {Ivanov},\ and\
			\citenamefont {Tobar}(2015)}]{bourhill2015}%
		\BibitemOpen
		\bibfield  {author} {\bibinfo {author} {\bibfnamefont {J.}~\bibnamefont
				{Bourhill}}, \bibinfo {author} {\bibfnamefont {E.}~\bibnamefont {Ivanov}}, \
			and\ \bibinfo {author} {\bibfnamefont {M.~E.}\ \bibnamefont {Tobar}},\
		}\bibfield  {title} {\enquote {\bibinfo {title} {Precision measurement of a
					low-loss cylindrical dumbbell-shaped sapphire mechanical oscillator using
					radiation pressure},}\ }\href@noop {} {\bibfield  {journal} {\bibinfo
				{journal} {Physical Review A}\ }\textbf {\bibinfo {volume} {92}},\ \bibinfo
			{pages} {023817} (\bibinfo {year} {2015})}\BibitemShut {NoStop}%
		\bibitem [{\citenamefont {Cotrufo}\ \emph {et~al.}(2018)\citenamefont
			{Cotrufo}, \citenamefont {Midolo}, \citenamefont {Zobenica}, \citenamefont
			{Petruzzella}, \citenamefont {Van~Otten},\ and\ \citenamefont
			{Fiore}}]{cotrufo2018}%
		\BibitemOpen
		\bibfield  {author} {\bibinfo {author} {\bibfnamefont {M.}~\bibnamefont
				{Cotrufo}}, \bibinfo {author} {\bibfnamefont {L.}~\bibnamefont {Midolo}},
			\bibinfo {author} {\bibfnamefont {Z.}~\bibnamefont {Zobenica}}, \bibinfo
			{author} {\bibfnamefont {M.}~\bibnamefont {Petruzzella}}, \bibinfo {author}
			{\bibfnamefont {F.~W.~M.}\ \bibnamefont {Van~Otten}}, \ and\ \bibinfo
			{author} {\bibfnamefont {A.}~\bibnamefont {Fiore}},\ }\bibfield  {title}
		{\enquote {\bibinfo {title} {Nanomechanical control of optical field and
					quality factor in photonic crystal structures},}\ }\href@noop {} {\bibfield
			{journal} {\bibinfo  {journal} {Physical Review B}\ }\textbf {\bibinfo
				{volume} {97}},\ \bibinfo {pages} {115304} (\bibinfo {year}
			{2018})}\BibitemShut {NoStop}%
		\bibitem [{\citenamefont {Abbott}\ \emph {et~al.}(2016)\citenamefont {Abbott},
			\citenamefont {Abbott}, \citenamefont {Abbott}, \citenamefont {Abernathy},
			\citenamefont {Acernese}, \citenamefont {Ackley}, \citenamefont {Adams},
			\citenamefont {Adams}, \citenamefont {Addesso}, \citenamefont {Adhikari},\
			and\ \citenamefont {et~al.}}]{abbott2016}%
		\BibitemOpen
		\bibfield  {author} {\bibinfo {author} {\bibfnamefont {B.~P.}\ \bibnamefont
				{Abbott}}, \bibinfo {author} {\bibfnamefont {R.}~\bibnamefont {Abbott}},
			\bibinfo {author} {\bibfnamefont {T.~D.}\ \bibnamefont {Abbott}}, \bibinfo
			{author} {\bibfnamefont {M.~R.}\ \bibnamefont {Abernathy}}, \bibinfo {author}
			{\bibfnamefont {F.}~\bibnamefont {Acernese}}, \bibinfo {author}
			{\bibfnamefont {K.}~\bibnamefont {Ackley}}, \bibinfo {author} {\bibfnamefont
				{C.}~\bibnamefont {Adams}}, \bibinfo {author} {\bibfnamefont
				{T.}~\bibnamefont {Adams}}, \bibinfo {author} {\bibfnamefont
				{P.}~\bibnamefont {Addesso}}, \bibinfo {author} {\bibfnamefont {R.~X.}\
				\bibnamefont {Adhikari}}, \ and\ \bibinfo {author} {\bibnamefont {et~al.}},\
		}\bibfield  {title} {\enquote {\bibinfo {title} {Observation of gravitational
					waves from a binary black hole merger},}\ }\href@noop {} {\bibfield
			{journal} {\bibinfo  {journal} {Physical review letters}\ }\textbf {\bibinfo
				{volume} {116}},\ \bibinfo {pages} {061102} (\bibinfo {year}
			{2016})}\BibitemShut {NoStop}%
		\bibitem [{\citenamefont {Bushev}\ \emph {et~al.}(2019)\citenamefont {Bushev},
			\citenamefont {Bourhill}, \citenamefont {Goryachev}, \citenamefont
			{Kukharchyk}, \citenamefont {Ivanov}, \citenamefont {Galliou}, \citenamefont
			{Tobar},\ and\ \citenamefont {Danilishin}}]{bushevQG}%
		\BibitemOpen
		\bibfield  {author} {\bibinfo {author} {\bibfnamefont {P.~A.}\ \bibnamefont
				{Bushev}}, \bibinfo {author} {\bibfnamefont {J.}~\bibnamefont {Bourhill}},
			\bibinfo {author} {\bibfnamefont {M.}~\bibnamefont {Goryachev}}, \bibinfo
			{author} {\bibfnamefont {N.}~\bibnamefont {Kukharchyk}}, \bibinfo {author}
			{\bibfnamefont {E.}~\bibnamefont {Ivanov}}, \bibinfo {author} {\bibfnamefont
				{S.}~\bibnamefont {Galliou}}, \bibinfo {author} {\bibfnamefont {M.~E.}\
				\bibnamefont {Tobar}}, \ and\ \bibinfo {author} {\bibfnamefont
				{S.}~\bibnamefont {Danilishin}},\ }\bibfield  {title} {\enquote {\bibinfo
				{title} {Testing of quantum gravity with sub-kilogram acoustic resonators},}\
		}\href@noop {} {\bibfield  {journal} {\bibinfo  {journal} {arXiv:1903.03346
					[quant-ph]}\ } (\bibinfo {year} {2019})}\BibitemShut {NoStop}%
		\bibitem [{\citenamefont {Teufel}\ \emph {et~al.}(2011)\citenamefont {Teufel},
			\citenamefont {Donner}, \citenamefont {Li}, \citenamefont {Harlow},
			\citenamefont {Allman}, \citenamefont {Cicak}, \citenamefont {Sirois},
			\citenamefont {Whittaker}, \citenamefont {Lehnert},\ and\ \citenamefont
			{Simmonds}}]{teufel2011}%
		\BibitemOpen
		\bibfield  {author} {\bibinfo {author} {\bibfnamefont {J.~D.}\ \bibnamefont
				{Teufel}}, \bibinfo {author} {\bibfnamefont {T.}~\bibnamefont {Donner}},
			\bibinfo {author} {\bibfnamefont {D.}~\bibnamefont {Li}}, \bibinfo {author}
			{\bibfnamefont {J.~W.}\ \bibnamefont {Harlow}}, \bibinfo {author}
			{\bibfnamefont {M.~S.}\ \bibnamefont {Allman}}, \bibinfo {author}
			{\bibfnamefont {K.}~\bibnamefont {Cicak}}, \bibinfo {author} {\bibfnamefont
				{A.~J.}\ \bibnamefont {Sirois}}, \bibinfo {author} {\bibfnamefont {J.~D.}\
				\bibnamefont {Whittaker}}, \bibinfo {author} {\bibfnamefont {K.~W.}\
				\bibnamefont {Lehnert}}, \ and\ \bibinfo {author} {\bibfnamefont {R.~W.}\
				\bibnamefont {Simmonds}},\ }\bibfield  {title} {\enquote {\bibinfo {title}
				{Sideband cooling of micromechanical motion to the quantum ground state},}\
		}\href@noop {} {\bibfield  {journal} {\bibinfo  {journal} {Nature}\ }\textbf
			{\bibinfo {volume} {475}},\ \bibinfo {pages} {359--363} (\bibinfo {year}
			{2011})}\BibitemShut {NoStop}%
		\bibitem [{\citenamefont {Weis}\ \emph {et~al.}(2010)\citenamefont {Weis},
			\citenamefont {Rivi{\`e}re}, \citenamefont {Del{\'e}glise}, \citenamefont
			{Gavartin}, \citenamefont {Arcizet}, \citenamefont {Schliesser},\ and\
			\citenamefont {Kippenberg}}]{weis2010}%
		\BibitemOpen
		\bibfield  {author} {\bibinfo {author} {\bibfnamefont {S.}~\bibnamefont
				{Weis}}, \bibinfo {author} {\bibfnamefont {R.}~\bibnamefont {Rivi{\`e}re}},
			\bibinfo {author} {\bibfnamefont {S.}~\bibnamefont {Del{\'e}glise}}, \bibinfo
			{author} {\bibfnamefont {E.}~\bibnamefont {Gavartin}}, \bibinfo {author}
			{\bibfnamefont {O.}~\bibnamefont {Arcizet}}, \bibinfo {author} {\bibfnamefont
				{A.}~\bibnamefont {Schliesser}}, \ and\ \bibinfo {author} {\bibfnamefont
				{T.~J.}\ \bibnamefont {Kippenberg}},\ }\bibfield  {title} {\enquote {\bibinfo
				{title} {Optomechanically induced transparency},}\ }\href@noop {} {\bibfield
			{journal} {\bibinfo  {journal} {Science}\ }\textbf {\bibinfo {volume}
				{330}},\ \bibinfo {pages} {1520--1523} (\bibinfo {year} {2010})}\BibitemShut
		{NoStop}%
		\bibitem [{\citenamefont {Renninger}\ \emph {et~al.}(2018)\citenamefont
			{Renninger}, \citenamefont {Kharel}, \citenamefont {Behunin},\ and\
			\citenamefont {Rakich}}]{renninger2018}%
		\BibitemOpen
		\bibfield  {author} {\bibinfo {author} {\bibfnamefont {W.~H.}\ \bibnamefont
				{Renninger}}, \bibinfo {author} {\bibfnamefont {P.}~\bibnamefont {Kharel}},
			\bibinfo {author} {\bibfnamefont {R.~O.}\ \bibnamefont {Behunin}}, \ and\
			\bibinfo {author} {\bibfnamefont {P.~T.}\ \bibnamefont {Rakich}},\ }\bibfield
		{title} {\enquote {\bibinfo {title} {Bulk crystalline optomechanics},}\
		}\href@noop {} {\bibfield  {journal} {\bibinfo  {journal} {Nature Physics}\
			}\textbf {\bibinfo {volume} {14}},\ \bibinfo {pages} {601} (\bibinfo {year}
			{2018})}\BibitemShut {NoStop}%
		\bibitem [{\citenamefont {Goryachev}\ and\ \citenamefont
			{Tobar}(2014)}]{goryachev2014e}%
		\BibitemOpen
		\bibfield  {author} {\bibinfo {author} {\bibfnamefont {M.}~\bibnamefont
				{Goryachev}}\ and\ \bibinfo {author} {\bibfnamefont {M.~E.}\ \bibnamefont
				{Tobar}},\ }\bibfield  {title} {\enquote {\bibinfo {title} {Effects of
					geometry on quantum fluctuations of phonon-trapping acoustic cavities},}\
		}\href@noop {} {\bibfield  {journal} {\bibinfo  {journal} {New Journal of
					Physics}\ }\textbf {\bibinfo {volume} {16}},\ \bibinfo {pages} {083007}
			(\bibinfo {year} {2014})}\BibitemShut {NoStop}%
		\bibitem [{\citenamefont {Galliou}\ \emph {et~al.}(2013)\citenamefont
			{Galliou}, \citenamefont {Goryachev}, \citenamefont {Bourquin}, \citenamefont
			{Abb{\'e}}, \citenamefont {Aubry},\ and\ \citenamefont
			{Tobar}}]{galliou2013}%
		\BibitemOpen
		\bibfield  {author} {\bibinfo {author} {\bibfnamefont {S.}~\bibnamefont
				{Galliou}}, \bibinfo {author} {\bibfnamefont {M.}~\bibnamefont {Goryachev}},
			\bibinfo {author} {\bibfnamefont {R.}~\bibnamefont {Bourquin}}, \bibinfo
			{author} {\bibfnamefont {P.}~\bibnamefont {Abb{\'e}}}, \bibinfo {author}
			{\bibfnamefont {J.~P.}\ \bibnamefont {Aubry}}, \ and\ \bibinfo {author}
			{\bibfnamefont {M.~E.}\ \bibnamefont {Tobar}},\ }\bibfield  {title} {\enquote
			{\bibinfo {title} {Extremely low loss phonon-trapping cryogenic acoustic
					cavities for future physical experiments},}\ }\href@noop {} {\bibfield
			{journal} {\bibinfo  {journal} {Scientific reports}\ }\textbf {\bibinfo
				{volume} {3}} (\bibinfo {year} {2013})}\BibitemShut {NoStop}%
		\bibitem [{\citenamefont {Le~Floch}\ \emph {et~al.}(2013)\citenamefont
			{Le~Floch}, \citenamefont {Fan}, \citenamefont {Aubourg}, \citenamefont
			{Cros}, \citenamefont {Carvalho}, \citenamefont {Shan}, \citenamefont
			{Bourhill}, \citenamefont {Ivanov}, \citenamefont {Humbert}, \citenamefont
			{Madrangeas},\ and\ \citenamefont {Tobar}}]{lefloch2013}%
		\BibitemOpen
		\bibfield  {author} {\bibinfo {author} {\bibfnamefont {J.-M.}\ \bibnamefont
				{Le~Floch}}, \bibinfo {author} {\bibfnamefont {Y.}~\bibnamefont {Fan}},
			\bibinfo {author} {\bibfnamefont {M.}~\bibnamefont {Aubourg}}, \bibinfo
			{author} {\bibfnamefont {D.}~\bibnamefont {Cros}}, \bibinfo {author}
			{\bibfnamefont {N.~C.}\ \bibnamefont {Carvalho}}, \bibinfo {author}
			{\bibfnamefont {Q.}~\bibnamefont {Shan}}, \bibinfo {author} {\bibfnamefont
				{J.}~\bibnamefont {Bourhill}}, \bibinfo {author} {\bibfnamefont {E.~N.}\
				\bibnamefont {Ivanov}}, \bibinfo {author} {\bibfnamefont {G.}~\bibnamefont
				{Humbert}}, \bibinfo {author} {\bibfnamefont {V.}~\bibnamefont {Madrangeas}},
			\ and\ \bibinfo {author} {\bibfnamefont {M.~E.}\ \bibnamefont {Tobar}},\
		}\bibfield  {title} {\enquote {\bibinfo {title} {Rigorous analysis of highly
					tunable cylindrical transverse magnetic mode re-entrant cavities},}\
		}\href@noop {} {\bibfield  {journal} {\bibinfo  {journal} {Review of
					Scientific Instruments}\ }\textbf {\bibinfo {volume} {84}},\ \bibinfo {pages}
			{125114} (\bibinfo {year} {2013})}\BibitemShut {NoStop}%
		\bibitem [{\citenamefont {Espinel}\ \emph {et~al.}(2017)\citenamefont
			{Espinel}, \citenamefont {Santos}, \citenamefont {Luiz}, \citenamefont
			{Alegre},\ and\ \citenamefont {Wiederhecker}}]{espinel2017}%
		\BibitemOpen
		\bibfield  {author} {\bibinfo {author} {\bibfnamefont {Y.~A.~V.}\
				\bibnamefont {Espinel}}, \bibinfo {author} {\bibfnamefont {F.~G.~S.}\
				\bibnamefont {Santos}}, \bibinfo {author} {\bibfnamefont {G.~O.}\
				\bibnamefont {Luiz}}, \bibinfo {author} {\bibfnamefont {T.~P.~M.}\
				\bibnamefont {Alegre}}, \ and\ \bibinfo {author} {\bibfnamefont {G.~S.}\
				\bibnamefont {Wiederhecker}},\ }\bibfield  {title} {\enquote {\bibinfo
				{title} {Brillouin optomechanics in coupled silicon microcavities},}\
		}\href@noop {} {\bibfield  {journal} {\bibinfo  {journal} {Scientific
					Reports}\ }\textbf {\bibinfo {volume} {7}} (\bibinfo {year}
			{2017})}\BibitemShut {NoStop}%
		\bibitem [{\citenamefont {Zou}\ \emph {et~al.}(2016)\citenamefont {Zou},
			\citenamefont {Han}, \citenamefont {Jiang},\ and\ \citenamefont
			{Tang}}]{zou2016}%
		\BibitemOpen
		\bibfield  {author} {\bibinfo {author} {\bibfnamefont {C.-L.}\ \bibnamefont
				{Zou}}, \bibinfo {author} {\bibfnamefont {X.}~\bibnamefont {Han}}, \bibinfo
			{author} {\bibfnamefont {L.}~\bibnamefont {Jiang}}, \ and\ \bibinfo {author}
			{\bibfnamefont {H.~X.}\ \bibnamefont {Tang}},\ }\bibfield  {title} {\enquote
			{\bibinfo {title} {Cavity piezomechanical strong coupling and frequency
					conversion on an aluminum nitride chip},}\ }\href@noop {} {\bibfield
			{journal} {\bibinfo  {journal} {Physical Review A}\ }\textbf {\bibinfo
				{volume} {94}},\ \bibinfo {pages} {013812} (\bibinfo {year}
			{2016})}\BibitemShut {NoStop}%
		\bibitem [{\citenamefont {Balram}\ \emph {et~al.}(2016)\citenamefont {Balram},
			\citenamefont {Davan{\c{c}}o}, \citenamefont {Song},\ and\ \citenamefont
			{Srinivasan}}]{balram2016}%
		\BibitemOpen
		\bibfield  {author} {\bibinfo {author} {\bibfnamefont {K.~C.}\ \bibnamefont
				{Balram}}, \bibinfo {author} {\bibfnamefont {M.~I.}\ \bibnamefont
				{Davan{\c{c}}o}}, \bibinfo {author} {\bibfnamefont {J.~D.}\ \bibnamefont
				{Song}}, \ and\ \bibinfo {author} {\bibfnamefont {K.}~\bibnamefont
				{Srinivasan}},\ }\bibfield  {title} {\enquote {\bibinfo {title} {Coherent
					coupling between radiofrequency, optical and acoustic waves in
					piezo-optomechanical circuits},}\ }\href@noop {} {\bibfield  {journal}
			{\bibinfo  {journal} {Nature photonics}\ }\textbf {\bibinfo {volume} {10}},\
			\bibinfo {pages} {346} (\bibinfo {year} {2016})}\BibitemShut {NoStop}%
		\bibitem [{\citenamefont {Bassan}\ \emph {et~al.}(2008)\citenamefont {Bassan},
			\citenamefont {Ballantini}, \citenamefont {Chincarini}, \citenamefont
			{Gemme}, \citenamefont {Iannuzzi}, \citenamefont {Moleti}, \citenamefont
			{Parodi},\ and\ \citenamefont {Vaccarone}}]{bassan2008}%
		\BibitemOpen
		\bibfield  {author} {\bibinfo {author} {\bibfnamefont {M.}~\bibnamefont
				{Bassan}}, \bibinfo {author} {\bibfnamefont {R.}~\bibnamefont {Ballantini}},
			\bibinfo {author} {\bibfnamefont {A.}~\bibnamefont {Chincarini}}, \bibinfo
			{author} {\bibfnamefont {G.}~\bibnamefont {Gemme}}, \bibinfo {author}
			{\bibfnamefont {M.}~\bibnamefont {Iannuzzi}}, \bibinfo {author}
			{\bibfnamefont {A.}~\bibnamefont {Moleti}}, \bibinfo {author} {\bibfnamefont
				{R.~F.}\ \bibnamefont {Parodi}}, \ and\ \bibinfo {author} {\bibfnamefont
				{R.}~\bibnamefont {Vaccarone}},\ }\bibfield  {title} {\enquote {\bibinfo
				{title} {New parametric transducer for resonant detectors: advances and room
					temperature test},}\ }\href@noop {} {\bibfield  {journal} {\bibinfo
				{journal} {Journal of Physics: Conference Series}\ }\textbf {\bibinfo
				{volume} {122}},\ \bibinfo {pages} {012031} (\bibinfo {year}
			{2008})}\BibitemShut {NoStop}%
		\bibitem [{\citenamefont {Vainsencher}\ \emph {et~al.}(2016)\citenamefont
			{Vainsencher}, \citenamefont {Satzinger}, \citenamefont {Peairs},\ and\
			\citenamefont {Cleland}}]{vainsencher2016}%
		\BibitemOpen
		\bibfield  {author} {\bibinfo {author} {\bibfnamefont {A.}~\bibnamefont
				{Vainsencher}}, \bibinfo {author} {\bibfnamefont {K.~J.}\ \bibnamefont
				{Satzinger}}, \bibinfo {author} {\bibfnamefont {G.~A.}\ \bibnamefont
				{Peairs}}, \ and\ \bibinfo {author} {\bibfnamefont {A.~N.}\ \bibnamefont
				{Cleland}},\ }\bibfield  {title} {\enquote {\bibinfo {title} {Bi-directional
					conversion between microwave and optical frequencies in a piezoelectric
					optomechanical device},}\ }\href@noop {} {\bibfield  {journal} {\bibinfo
				{journal} {Applied Physics Letters}\ }\textbf {\bibinfo {volume} {109}},\
			\bibinfo {pages} {033107} (\bibinfo {year} {2016})}\BibitemShut {NoStop}%
		\bibitem [{\citenamefont {Javerzac-Galy}\ \emph {et~al.}(2016)\citenamefont
			{Javerzac-Galy}, \citenamefont {Plekhanov}, \citenamefont {Bernier},
			\citenamefont {Toth}, \citenamefont {Feofanov},\ and\ \citenamefont
			{Kippenberg}}]{javerzac2016}%
		\BibitemOpen
		\bibfield  {author} {\bibinfo {author} {\bibfnamefont {C.}~\bibnamefont
				{Javerzac-Galy}}, \bibinfo {author} {\bibfnamefont {K.}~\bibnamefont
				{Plekhanov}}, \bibinfo {author} {\bibfnamefont {N.~R.}\ \bibnamefont
				{Bernier}}, \bibinfo {author} {\bibfnamefont {L.~D.}\ \bibnamefont {Toth}},
			\bibinfo {author} {\bibfnamefont {A.~K.}\ \bibnamefont {Feofanov}}, \ and\
			\bibinfo {author} {\bibfnamefont {T.~J.}\ \bibnamefont {Kippenberg}},\
		}\bibfield  {title} {\enquote {\bibinfo {title} {On-chip microwave-to-optical
					quantum coherent converter based on a superconducting resonator coupled to an
					electro-optic microresonator},}\ }\href@noop {} {\bibfield  {journal}
			{\bibinfo  {journal} {Physical Review A}\ }\textbf {\bibinfo {volume} {94}},\
			\bibinfo {pages} {053815} (\bibinfo {year} {2016})}\BibitemShut {NoStop}%
		\bibitem [{\citenamefont {Han}, \citenamefont {Zou},\ and\ \citenamefont
			{Tang}(2016)}]{han2016}%
		\BibitemOpen
		\bibfield  {author} {\bibinfo {author} {\bibfnamefont {X.}~\bibnamefont
				{Han}}, \bibinfo {author} {\bibfnamefont {C.-L.}\ \bibnamefont {Zou}}, \ and\
			\bibinfo {author} {\bibfnamefont {H.~X.}\ \bibnamefont {Tang}},\ }\bibfield
		{title} {\enquote {\bibinfo {title} {Multimode strong coupling in
					superconducting cavity piezoelectromechanics},}\ }\href@noop {} {\bibfield
			{journal} {\bibinfo  {journal} {Physical review letters}\ }\textbf {\bibinfo
				{volume} {117}},\ \bibinfo {pages} {123603} (\bibinfo {year}
			{2016})}\BibitemShut {NoStop}%
		\bibitem [{\citenamefont {Carvalho}, \citenamefont {Fan},\ and\ \citenamefont
			{Tobar}(2016)}]{carvalho2016}%
		\BibitemOpen
		\bibfield  {author} {\bibinfo {author} {\bibfnamefont {N.~C.}\ \bibnamefont
				{Carvalho}}, \bibinfo {author} {\bibfnamefont {Y.}~\bibnamefont {Fan}}, \
			and\ \bibinfo {author} {\bibfnamefont {M.~E.}\ \bibnamefont {Tobar}},\
		}\bibfield  {title} {\enquote {\bibinfo {title} {Piezoelectric tunable
					microwave superconducting cavity},}\ }\href@noop {} {\bibfield  {journal}
			{\bibinfo  {journal} {Review of Scientific Instruments}\ }\textbf {\bibinfo
				{volume} {87}},\ \bibinfo {pages} {094702} (\bibinfo {year}
			{2016})}\BibitemShut {NoStop}%
		\bibitem [{\citenamefont {Goryachev}\ \emph {et~al.}(2013)\citenamefont
			{Goryachev}, \citenamefont {Creedon}, \citenamefont {Galliou},\ and\
			\citenamefont {Tobar}}]{goryachev2013o}%
		\BibitemOpen
		\bibfield  {author} {\bibinfo {author} {\bibfnamefont {M.}~\bibnamefont
				{Goryachev}}, \bibinfo {author} {\bibfnamefont {D.~L.}\ \bibnamefont
				{Creedon}}, \bibinfo {author} {\bibfnamefont {S.}~\bibnamefont {Galliou}}, \
			and\ \bibinfo {author} {\bibfnamefont {M.~E.}\ \bibnamefont {Tobar}},\
		}\bibfield  {title} {\enquote {\bibinfo {title} {Observation of rayleigh
					phonon scattering through excitation of extremely high overtones in low-loss
					cryogenic acoustic cavities for hybrid quantum systems},}\ }\href@noop {}
		{\bibfield  {journal} {\bibinfo  {journal} {Physical Review Letters}\
			}\textbf {\bibinfo {volume} {111}},\ \bibinfo {pages} {085502} (\bibinfo
			{year} {2013})}\BibitemShut {NoStop}%
	\end{thebibliography}
	%

\end{document}